\documentclass[aps,prl,twocolumn,groupedaddress,showpacs,showkeys, 10pt]{revtex4-1}
\pdfoutput=1
\usepackage{graphicx}
\usepackage{siunitx}

\DeclareSIUnit\erec{\text{\ensuremath {\text{E}_{\textup {rec}}}}}
\DeclareSIUnit\deg{\degree}
\DeclareSIUnit\nph{\text{nph}}
\DeclareSIUnit\cts{\text{Cts}}

\def \beq{\begin{equation}}
\def \eeq{\end{equation}}
\def \bea{\begin{eqnarray}}
\def \eea{\end{eqnarray}}
\def \bem{\begin{displaymath}}
\def \eem{\end{displaymath}}

\DeclareSIUnit[]\Er
{\text{\ensuremath {E_{\mathrm{R}}}}}

\begin{document}

\title{Dissipation Induced Structural Instability and Chiral Dynamics in a Quantum Gas}

\author{Nishant Dogra}
\affiliation{Institute for Quantum Electronics, ETH Zurich, CH-8093 Zurich, Switzerland}

\author{Manuele Landini}
\affiliation{Institute for Quantum Electronics, ETH Zurich, CH-8093 Zurich, Switzerland}

\author{Katrin Kroeger}
\affiliation{Institute for Quantum Electronics, ETH Zurich, CH-8093 Zurich, Switzerland}

\author{Lorenz Hruby}
\affiliation{Institute for Quantum Electronics, ETH Zurich, CH-8093 Zurich, Switzerland}

\author{Tobias Donner}
\email[]{donner@phys.ethz.ch}
\affiliation{Institute for Quantum Electronics, ETH Zurich, CH-8093 Zurich, Switzerland}

\author{Tilman Esslinger}
\affiliation{Institute for Quantum Electronics, ETH Zurich, CH-8093 Zurich, Switzerland}

\bibliographystyle{apsrev4-1}

\date{\today}

\def \beq{\begin{equation}}
\def \eeq{\end{equation}}
\def \bea{\begin{eqnarray}}
\def \eea{\end{eqnarray}}
\newcommand{\avg}[1] {
	\langle #1 \rangle}

\begin{abstract}
Dissipative and unitary processes define the evolution of a many-body system. Their interplay gives rise to dynamical phase transitions and can lead to instabilities. We discovered a non-stationary state of chiral nature in a synthetic many-body system with independently controllable unitary and dissipative couplings. Our experiment is based on a spinor Bose gas interacting with an optical resonator. Orthogonal quadratures of the resonator field coherently couple the Bose-Einstein condensate to two different atomic spatial modes whereas the dispersive effect of the resonator losses mediates a dissipative coupling between these modes. In a regime of dominant dissipative coupling we observe the chiral evolution and map it to a positional instability.
\end{abstract}

\maketitle

In a many-body system, unitary processes give rise to coherent evolution, while dissipative processes lead to stationarity \cite{Breuer2007}. Also the interplay of these processes in a driven-dissipative setting can influence a many-body system in profound ways. Examples are dissipative phase transitions \cite{Syassen2008,Barontini2013,Brennecke2013,Labouvie2016,Tomita2017,Carusotto2013,Fink2018}, the emergence of new universality classes \cite{Diehl2010,Nagy2011}, dissipation-induced topological effects \cite{Diehl2011}, complex dynamics \cite{Buca2018}, or the splitting of multi-critical points \cite{Soriente2018}. Here we discover a phenomenon where a chiral non-stationary dynamics emerges if the energy scales of dissipative and unitary processes are similar. In our experiment, we create a driven many-body system with controllable unitary and dissipative couplings using a quantum gas. This allows us to explore the system's macroscopic behavior at the boundary between stationary and non-stationary states. We gain a conceptual understanding of the observed dynamics by considering dissipation as a structure dependent  force, in close analogy to mechanical non-conservative positional forces \cite{Newkirk1925,Kapitza1939,Merkin1996}.

\begin{figure}[ht]
	\includegraphics[width=\columnwidth]{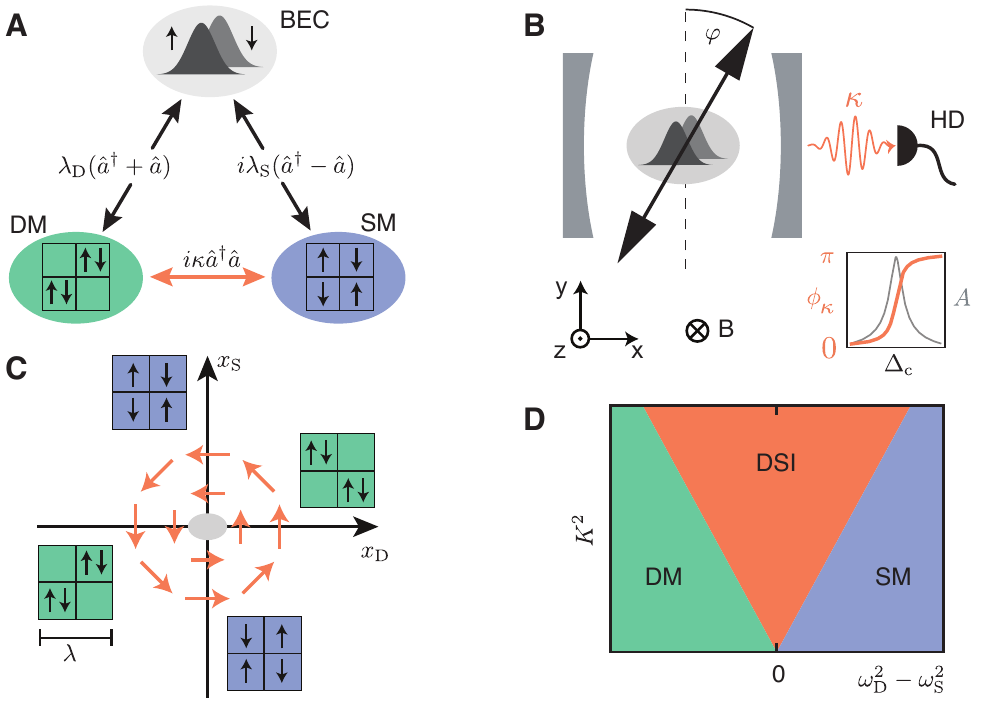}
	\caption{\textbf{Engineered dissipative coupling:} (\textbf{A}) A spinor BEC containing a mixture of two Zeeman states (depicted by up and down arrows) is coherently coupled with rates $\lambda_\mathrm{D, S}$ to a density (DM) and a spin (SM) mode via the two quadratures of a cavity field with annihilation operator $\hat{a}$. A dissipative coupling (orange arrow) between DM and SM is induced by the phase response of the cavity. Occupation of DM (SM) leads to a density (spin) modulation, displayed as atoms occupying the same (opposite) checkerboard lattice sites of an exemplary green (blue) 2 $\times$ 2 square. (\textbf{B}) The spinor BEC is coupled with a cavity mode and irradiated by a standing wave transverse pump in z-direction with linear polarization (black arrow) at angle $\varphi$ with respect to the y-direction. The magnetic field B is oriented along the z-direction. Light leaking out of the cavity at decay rate $\kappa$ is analyzed via a heterodyne detector HD. Inset: phase $\phi_\kappa$ and amplitude $A$ response of the cavity as a function of detuning $\Delta_{\mathrm{c}}$.  (\textbf{C}) Dissipative chiral force (orange) where $x_{\mathrm{D(S)}}$ represents the amplitude of the density (spin) modulation. (\textbf{D}) Three different regimes of the system as function of dissipative coupling strength $K^2$ and detuning between the frequencies $\omega_{\mathrm{D},\mathrm{S}}$ of the DM and SM. Green (blue) region represents the DM (SM) being dominantly coupled to the spinor BEC. The dissipation induced structural instability (DSI) is depicted in orange.}
	\label{fig:Fig1}
\end{figure}

Our experiment consists of a spinor Bose-Einstein condensate (BEC) of two different Zeeman states that is coherently coupled to two different spatial atomic configurations \cite{Landini2018}, referred to as density mode (DM) and spin mode (SM), Fig. \ref{fig:Fig1}. These coherent couplings are mediated via photons scattered by the atomic system from a standing wave transverse pump laser field into a high finesse optical cavity mode. The DM and the SM interact with orthogonal quadratures of the cavity mode. We engineer a dissipative coupling in this system exploiting the finite cavity decay rate $\kappa$ and the associated phase shift of the intra-cavity field across the cavity resonance: the light field scattered from the pump into the cavity acquires a phase shift $\phi_\kappa$ that effectively mixes the orthogonal quadratures, giving rise to a dissipative coupling between the DM and the SM, Fig. \ref{fig:Fig1}.

The strengths $\lambda_{\mathrm{D},\mathrm{S}}$ of the coherent couplings between the BEC and the DM or the SM, respectively, are tuned by the lattice depth $V_\mathrm{TP}$ and polarization angle $\varphi$ of the transverse pump \cite{Landini2018}. These couplings soften the effective excitation frequencies $\omega_{\mathrm{D},\mathrm{S}}$ of both modes, such that at a critical lattice depth the frequency of the more strongly coupled mode vanishes. This mode can then be macroscopically occupied, and the system undergoes a self-organization phase transition \cite{Baumann2010}, breaking a spatial Z(2) symmetry. Simultaneously, the corresponding quadrature of the cavity mode is coherently populated, which we detect with a heterodyne detection system analyzing the light field leaking from the cavity \cite{Landig2015}.

\begin{figure*}
	\includegraphics[]{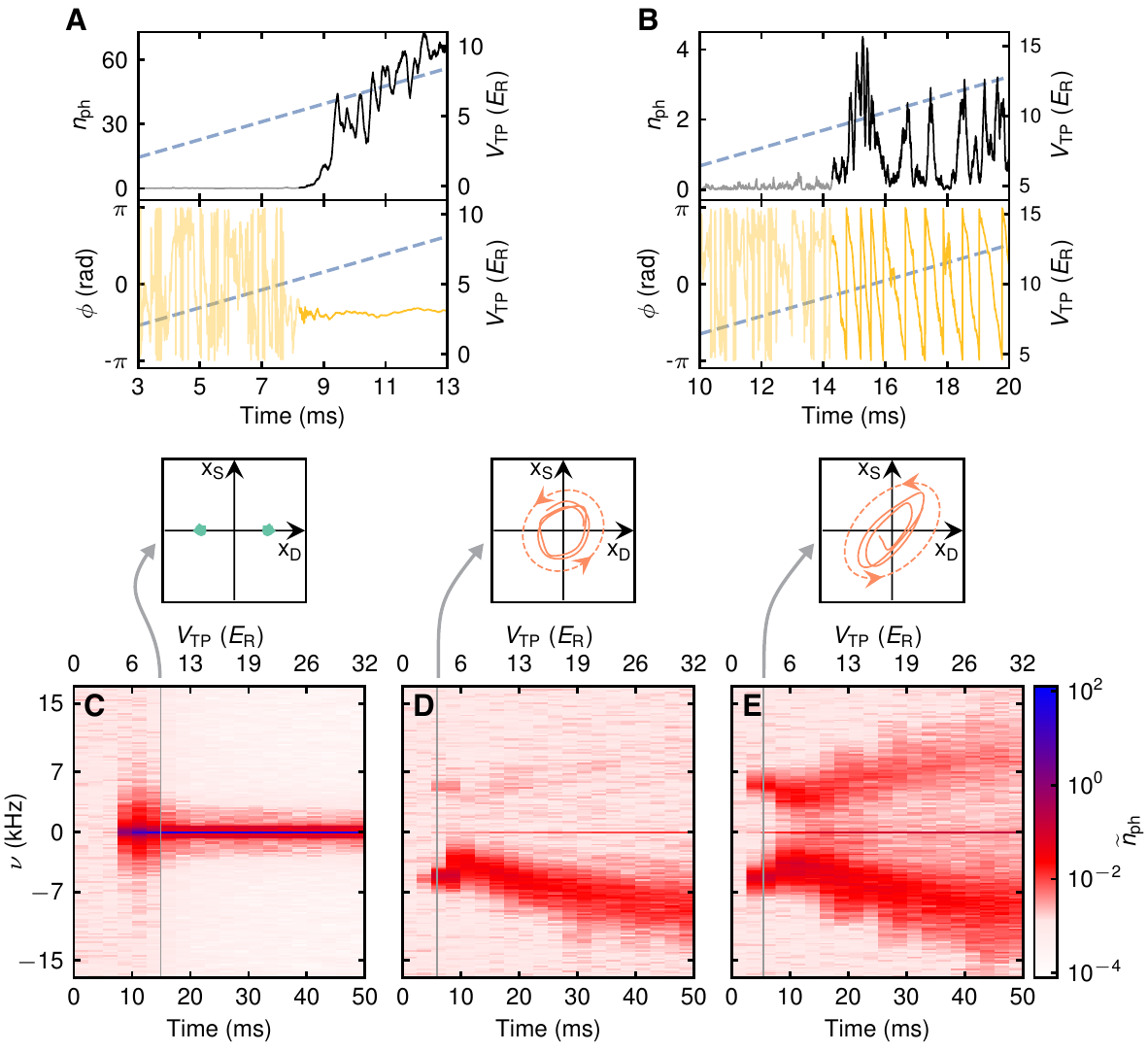}%
	\caption{\textbf{Detection of dissipation induced structural instability:} Time evolution of the mean number of photons $n_{\mathrm{ph}}$ (black) in the cavity and the corresponding phase $\phi$ (orange) for (\textbf{A}) $\varphi = 58.3(1.0)^{\circ}$, $\tilde{\Delta}_{\mathrm{c}}/2\pi = -5.5(1)\,\mathrm{MHz}$ and (\textbf{B}) $\varphi = 63.3(1.0)^{\circ}$, $\tilde{\Delta}_{\mathrm{c}}/2\pi = -5.5(1)\,\mathrm{MHz}$. The detuning $\tilde{\Delta}_{\mathrm{c}}$ includes the effect of the atomic dispersive shift on the detuning $\Delta_{\mathrm{c}}$ (see SI). The linear ramp of the lattice depth $V_{\mathrm{TP}}$ of the transverse pump is shown as dashed blue line. Grey and light orange lines correspond to zero occupation of the cavity. Spectrograms showing the mean number of photons $\tilde{n}_{\mathrm{ph}}$ as a function of frequency $\nu$ and time for (\textbf{C}) $\varphi = 58.3(1.0)^{\circ}$, $\tilde{\Delta}_{\mathrm{c}}/2\pi = -5.9(3)\,\mathrm{MHz}$, (\textbf{D}) $\varphi = 63.3(1.0)^{\circ}$, $\tilde{\Delta}_{\mathrm{c}}/2\pi = -2.9(3)\,\mathrm{MHz}$ and (\textbf{E}) $\varphi = 58.3(1.0)^{\circ}$, $\tilde{\Delta}_{\mathrm{c}}/2\pi = -2.9(3)\,\mathrm{MHz}$. A window size of 5\,ms is used to construct the spectrograms and each data set is averaged over 20 experimental realizations. Frequency $\nu$ is defined relative to the transverse pump. Corresponding sub-panels show the evolution of the atomic state in the $x_{\mathrm{D}}-x_{\mathrm{S}}$ plane extracted from the spectrograms in a duration of 500\,$\mu$s at the position of the gray lines. In the sub-panel associated with (\textbf{C}), we plot the data and its mirror about $x_{\mathrm{S}}$ to illustrate the Z(2) symmetry breaking of the self-organization phase transition. Dashed lines in the sub-panels corresponding to (\textbf{D}) and (\textbf{E}) are the predictions of the non-interacting theory. All the sub-panels are rescaled with respect to each other and the theoretical lines for better visibility.}
	\label{fig:Fig2}	
\end{figure*}

The cavity induced phase shift $\phi_\kappa$ and hence the dissipative coupling strength $K^2$ between the two modes can be controlled by the detuning $\Delta_{\mathrm{c}}$ between the cavity resonance and the frequency of the transverse pump (see SI). The effect of this dissipative coupling can be understood in the $x_{\mathrm{D}}-x_{\mathrm{S}}$ plane, where $x_{\mathrm{D},\mathrm{S}}$ represent the average amplitudes of the density and spin modulation caused by the occupation of DM and SM, respectively,  Fig. \ref{fig:Fig1}C. In this plane, the dissipative coupling acts as a force field that favors a rotation of the system’s state around the origin. If the DM and the SM are degenerate, already an infinitesimally small dissipative coupling leads to a structural instability (DSI) where the system rotates with fixed chirality between the different atomic modes. Increasing the strength of the dissipative coupling is expected to broaden the region of instability as shown schematically in Fig. \ref{fig:Fig1}D.

We prepare the $^{87}$Rb spinor BEC with $N = (23 \pm 2)\times 10^3$ atoms in each of the different Zeeman states  $|F=1, m_F = \pm 1 \rangle$, where $F$ and $m_F$ denote the total angular momentum and the corresponding magnetic quantum number. We linearly ramp up the lattice depth of the transverse pump in 50\,ms and analyze the cavity output with our heterodyne setup. In Fig. \ref{fig:Fig2}A-B, we show the mean photon number $n_{\mathrm{ph}}$ and phase $\phi$ (modulo $2\pi$) of the intra-cavity light field for two different sets of parameters. We find two qualitatively different behaviors: above a critical pump power, the cavity field has a non-zero amplitude and either a well-defined (Fig. \ref{fig:Fig2}A), or a monotonically changing phase (Fig. \ref{fig:Fig2}B). A well-defined phase indicates that only one quadrature of the cavity field is excited, corresponding to either the SM or the DM being populated, decided by which coherent coupling prevails. In contrast, a monotonically changing phase is observed when the dissipative coupling is dominant, and signals that the system is continuously evolving through the different spatial modes linked with the two quadratures of the cavity field.

The many-body Hamiltonian of the system consists of terms describing the energies of the bare atomic and photonic modes, and a term capturing the couplings of the DM and the SM to the respective quadratures of the cavity field  (see SI). The damping of the cavity mode can be modeled by a non-Hermitian term $-i\kappa \hat{a}^{\dagger}\hat{a}$ in the Hamiltonian, where $\hat{a}$ is the annihilation operator of the cavity field in a frame rotating at the transverse pump frequency. It causes a phase shift $\phi_{\kappa} = \tan^{-1}(-\kappa/\Delta_{\mathrm{c}})$ of the field scattered into the cavity by the atomic system. We adiabatically eliminate the cavity field and write the linearized equations of motion near the ground state of the non-interacting spinor BEC for the average amplitudes $x_{\mathrm{D}}$ and $x_{\mathrm{S}}$ (see SI):
\beq
\frac{d^2}{dt^2}
\left(
\begin{array}{c}
	x_{\mathrm{D}}\\
	x_{\mathrm{S}}\\
\end{array}
\right)
= \left(
\begin{array}{cc}
	-\omega_{\mathrm{D}}^2 & -K^2 \\
	K^2 & -\omega_{\mathrm{S}}^2 \\
\end{array}
\right)
\left(
\begin{array}{c}
	x_{\mathrm{D}}\\
	x_{\mathrm{S}}\\
\end{array}
\right)\,,
\label{eq:dissipatively coupled modes}
\eeq
where $\omega^2_{\mathrm{D}} - \omega^2_{\mathrm{S}} \propto \sin{\delta\phi}$ with $\delta\phi = 2\tan^{-1}(\lambda_{\mathrm{S}}/\lambda_{\mathrm{D}}) - \pi/2$ is a measure of the relative coupling strength of the BEC to the SM and the DM. The strength of the dissipative coupling $K^2 \propto V_{\mathrm{TP}} \sin^2{\phi_{\kappa}}\cos{\delta\phi}$ can thus be enhanced by either increasing the cavity induced phase shift or by making the two modes degenerate. This dissipative coupling generates a chiral force orthogonal to the current position vector of the system in the $x_{\mathrm{D}}-x_{\mathrm{S}}$ plane, Fig. \ref{fig:Fig1}C, and provides an example of a positional force \cite{Merkin1996,Krechetnikov2007}. Such positional forces are known in mechanical systems like a rotating shaft subject to friction due to an incompressible viscous fluid in a bearing. The incompressibility of the fluid leads to unequal frictional forces on the opposite sides of the shaft, resulting in a positional force orthogonal to the direction of the displacement \cite{Newkirk1925,Kapitza1939}. In our system, when the two atomic modes are degenerate, this positional force cannot be counteracted by the restoring harmonic force which is pointing towards the origin, leading to a dissipation induced structural or positional instability \cite{Merkin1996, 1982rotordynamic,Krechetnikov2007}.

Mode degeneracy (i.e. $\lambda_\mathrm{D}=\lambda_\mathrm{S}$ such that $\delta \phi = 0$) is reached at the critical polarization angle $\varphi_{\mathrm{c}} = 65.1^\circ$ for the chosen wavelength of the transverse pump of $\lambda = 784.7\, \mathrm{nm}$. To explain our observations, we analyze the solutions of Eq. \ref{eq:dissipatively coupled modes} and the corresponding intra-cavity light field for polarization angles close to $\varphi_{\mathrm{c}}$, that is, $\delta \phi$ being small (see SI): 
\begin{eqnarray}
\big[x_{\mathrm{D}}^{\mathrm{c}},x_{\mathrm{S}}^{\mathrm{c}}\big] &=& A\big[ \cos{\omega t},\sin{(\omega t + \frac{\Delta_{\mathrm{c}}}{\kappa}\delta \phi)} \big] e^{gt}  \nonumber \\ %
\avg{\hat{a}^{\dagger}} &\propto& (x_{\mathrm{D}}^{\mathrm{c}} -ix_{\mathrm{S}}^{\mathrm{c}}) - \frac{\delta \phi}{2} (x_{\mathrm{D}}^{\mathrm{c}} + ix_{\mathrm{S}}^{\mathrm{c}}) \nonumber \\
&\propto& A\Big[ e^{-i\omega t}  - \frac{\delta \phi}{2} (1 + i\frac{\Delta_c}{\kappa}) e^{i\omega t} \Big] e^{gt}\,.
\label{eq:elliptical chiral state}
\end{eqnarray}
In the limit $\delta \phi = 0$,  such a time-dependent solution implies that the system is rotating in the $x_{\mathrm{D}}-x_{\mathrm{S}}$ plane with fixed chirality at frequency $\omega$ and amplification rate $g$ (see SI). Microscopically, this rotation is associated with the atomic spins moving from one $\lambda$-periodic spatial pattern to another, Fig. \ref{fig:Fig1}C. Since the two atomic modes are connected to different quadratures of the cavity, the phase of the cavity field evolves monotonically as observed in Fig. \ref{fig:Fig2}B and shown in Eq. \ref{eq:elliptical chiral state}.

The frequency spectrum of the light leaking from the cavity is also accessible with our heterodyne setup \cite{Landig2015}. Fig. \ref{fig:Fig2}C-E show spectrograms for three different sets of parameters of data similar to Fig. \ref{fig:Fig2}A-B, but averaged over 20 repetitions. Fig. \ref{fig:Fig2}C shows a spectrogram where the signal is located at zero frequency ($\nu = 0$). It corresponds to the frequency of the transverse pump and is identical to the observation of a constant time phase of the cavity field as shown in Fig. \ref{fig:Fig2}A. We identify this as the formation of a static checkerboard density pattern which coherently scatters the pump field into the cavity \cite{Landini2018} (see SI). The corresponding steady state of the system is displayed in the sub-panel of Fig. \ref{fig:Fig2}C.

In contrast, Fig. \ref{fig:Fig2}D-E depict red ($\nu < 0$) and blue ($\nu > 0$) detuned sidebands with the peak frequency being a function of the lattice depth of the transverse pump. Observation of only a red sideband, as in  Fig. \ref{fig:Fig2}D, is the counterpart of a linearly running phase. For small lattice depths ($V_{\mathrm{TP}} < 6\, E_\mathrm{R}$), the sideband frequency is expected to be the root mean square of the two mode frequencies, that is, $\omega \approx \omega_{\mathrm{e}} = \sqrt{(\omega_{\mathrm{D}}^2 + \omega_{\mathrm{S}}^2)/2}$. Evolution at this intermediate frequency reflects a synchronization process \cite{Pikovsky2002} between the two spatial modes arising from the dissipative coupling. For large lattice depths, the sideband frequency depends on the dissipative coupling strength: $\omega \approx K^2/2|\omega_e|$ (see SI). In this limit, the two mode frequencies become imaginary which would correspond to the self-organization phase transition in the absence of dissipative coupling.

The relative strength $R$ of the blue with respect to the red sideband increases towards one as $\varphi$ deviates from the critical angle $\varphi_\mathrm{c}$. The presence of the blue sideband is connected to non-zero $\delta \phi$ (Eq. \ref{eq:elliptical chiral state}) and leads to an elliptical evolution in the $x_{\mathrm{D}} - x_{\mathrm{S}}$ plane. The relative strength $R$, and hence the ellipticity of the chiral solution can be influenced via $\Delta_\mathrm{c}$ or $|\delta \phi|$. Microscopically, the blue sideband is connected to the motion of a different number of atoms in each Zeeman state. The sub-panels of Fig. \ref{fig:Fig2}D-E show data of the time varying trajectory of the system together with solutions obtained from Eq. \ref{eq:elliptical chiral state} for $g = 0$, illustrating the non-stationary chiral state. The number of photons observed in the case of the DSI (Fig. \ref{fig:Fig2}D-E) is much smaller than during self-organization (Fig. \ref{fig:Fig2}C), although the instability is associated with a finite amplification rate $g$. We attribute this, and the observed pulsing behavior in the number of photons (Fig. \ref{fig:Fig2}B), to collisional interactions between the atoms (see SI).

\begin{figure}
	\includegraphics[keepaspectratio=true]{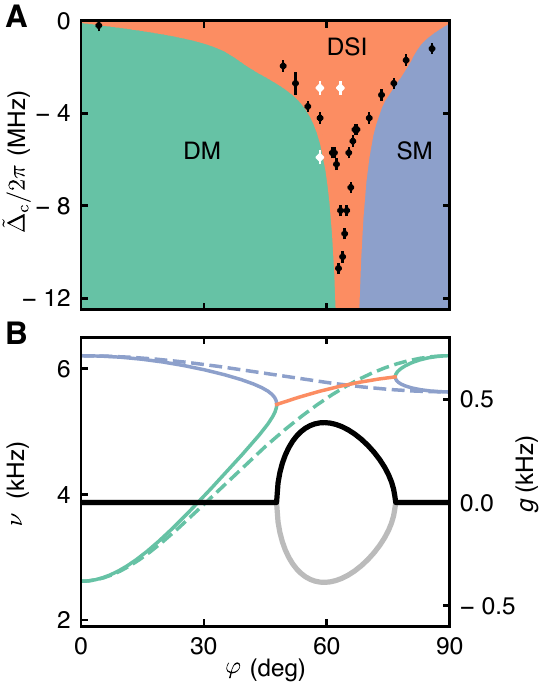}%
	\caption{\textbf{Boundary of the instability dominated regime:} (\textbf{A}) Critical detuning, including the dispersive shift (see SI), for the onset of the dissipation induced instability as a function of polarization angle $\varphi$. The black data points indicate the smallest detuning where the strength of the sidebands becomes larger than the signal at the transverse pump frequency in the spectrograms for a fixed lattice depth of $21\,E_\mathrm{R}$. The white diamonds indicate the parameters for the spectrograms shown in Fig. \ref{fig:Fig2}C-E. Errorbars are given by the discretized detuning interval in which the transition was detected. The polarization angle determination has an error of $1^{\circ}$. The green and blue regions correspond to DM and SM being respectively more strongly coupled to the spinor BEC. In the orange region, the system is dominated by the dissipation induced instability. The boundary between different regions is obtained from a non-interacting theory, see main text and (see SI). (\textbf{B}) Green and blue solid (dashed) lines show the frequency $\nu$ of the two eigenmodes of the system as a function of the polarization angle in the presence (absence) of dissipative coupling. Without dissipation, the two modes are the DM (green) and the SM (blue). Dissipation leads to level attraction between the two modes and hence changes their frequencies. In the instability dominated regime, the two modes are synchronized with each other (orange line). Correspondingly, the gain $g$ of the amplified (black line) and the damped mode (gray line) are plotted. The lines are drawn for $\tilde{\Delta}_{\mathrm{c}}/2\pi = -2.7\,\mathrm{MHz}$ or $\Delta_{\mathrm{c}}/2\pi = -4\,\mathrm{MHz}$ and $V_{\mathrm{TP}} = 1.5 E_\mathrm{R}$.}
	\label{fig:Fig3}
\end{figure}

We experimentally map out the boundary of the DSI region by ramping up the transverse pump lattice to 25\,$E_\mathrm{R}$ in 50\,ms for various polarization angles $\varphi$ and detunings $\Delta_{\mathrm{c}}$, Fig. \ref{fig:Fig3}A. Here, $E_\mathrm{R}=\hbar^2 k^2/2m$ is the  recoil energy with wave vector $k$ of the transverse pump and mass $m$ of a $^{87}$Rb atom, and $2\pi \hbar$ is Planck's constant. For a given polarization angle, we define the onset of the DSI by the minimum detuning where the strength of the sidebands exceeds the signal at the transverse pump frequency.  We theoretically obtain the boundary of the DSI (orange region in \ref{fig:Fig3}A) from Eq. \ref{eq:dissipatively coupled modes} as $|\omega_{\mathrm{D}}^2 - \omega_{\mathrm{S}}^2| = 2K^2$ which results in a critical detuning $\Delta_{\mathrm{c},\mathrm{b}}= -\kappa / \tan{|\delta\phi|}$.  A study of the extent of the DSI as a function of polarization angle and transverse pump lattice depth is presented in the supplementary information.

Physically, the boundary of the DSI can be understood as an onset of synchronization between DM and SM. Such a synchronization process is a result of a dissipation induced level attraction, a hallmark of non-Hermitian systems \cite{Rotter2009,El-Ganainy2018}. Fig. \ref{fig:Fig3}B shows the frequencies of the DM and SM in the absence (dashed lines) and presence (solid lines) of dissipation. In the vicinity of the critical angle, this level attraction leads to the emergence of two degenerate modes with opposite chirality. While one of them is damped, the other mode is amplified which gives rise to the DSI.

We have experimentally studied a many-body system where both coherent and dissipative couplings are independently tunable. Our observation of a dissipation induced chiral instability represents a new form of quantum many-body dynamics such as limit cycles \cite{Gutzwiller1990,Keeling2010a,Piazza2015} and time crystals \cite{Zhang2017,Choi2017}.


\section*{Acknowledgments}
We thank Berislav Buca, Ezequiel Rodriguez Chiacchio, Francesco Ferri, Dieter Jaksch, Andreas Nunnenkamp, and Joseph Tindall for stimulating discussions. We acknowledge funding from SNF: project numbers 182650 and175329 (NAQUAS QuantERA) and NCCR QSIT, from EU Horizon2020: ERCadvanced grant TransQ (Project Number 742579), from SBFI (QUIC, contract No. 15.0019).

\onecolumngrid
\section*{Supplementary Information}
\subsection*{Experimental Details}
We prepare an atomic cloud with either \num{45 \pm 1 e3} atoms in state $|F=1,m_F=0\rangle$ or \num{23 \pm 2 e3} atoms in each $|F=1,m_F=\pm 1\rangle$ spin state, where $F$ and $m_F$ represent the total angular momentum and the corresponding magnetic quantum number. The quantization axis is defined by a magnetic field of about 137 Gauss pointing in the negative z-direction which results in a Zeeman splitting of $\Delta E/\hbar \approx 2\pi \times\SI{96}{\mega\hertz}$. The trapping frequencies of the crossed dipole trap which holds the atoms at the center of the cavity mode amount to $\omega_{x,y,z}/2\pi=[329(6),62(3),168(2)] \si{\hertz}$. The lattice depth of the transverse pump is calibrated with Raman-Nath diffraction \cite{Morsch2006}. Details of the experimental techniques, in particular spin preparation, spin changing methods and polarization control, can be found in \cite{Landini2018}.

\subsection*{Atom number determination}
The atom numbers in the different $m_F$ states are determined from absorption images after ballistic expansion with a Stern-Gerlach measurement. Due to spurious magnetic gradients during the imaging sequence, the resonance frequency is inhomogeneous across the imaging region, leading to different detection efficiencies for $m_F=\pm1$. We accounted for this inhomogeneous detection efficiency in the data evaluation.\\
For measurements performed on a spin mixture, we use absorption images recorded at the end of each experimental run to discard runs which had an inefficient transfer from the state $m_F=0$ to the spin mixture. We required that not more than 15\si{\percent} of the atoms remained in state $m_F=0$.\\

\subsection*{Measurement protocols}
In order to measure the spectrograms (Fig. \ref{fig:Fig2}), the transverse pump power is ramped up linearly in \SI{50}{\milli\second}, held for \SI{50}{\milli\second}, and ramped down linearly in \SI{50}{\milli\second}.

For measuring the boundary of the instability dominated regime (Fig. \ref{fig:Fig3}), the transverse pump power is ramped up via an S-shaped ramp during \SI{50}{\milli\second}, held for \SI{50}{\milli\second}, and ramped down again via the S-shaped ramp in \SI{50}{\milli\second}. The S-shaped ramp has the form: $V(t) = V_0\Big[3\Big( \frac{t}{t_0}\Big)^2 - 2\Big( \frac{t}{t_0}\Big)^3 \Big]$. Here $V_0$ is the final lattice depth, $t$ is the time and $t_0$ is the full duration of the ramp.

For the measurement of the phase of the cavity light field to determine whether the DM or SM is populated \cite{Landini2018}, the transverse pump power is twice ramped up via the S-shaped ramp, held, and ramped down via the S-shaped ramp for \SI{50}{\milli\second} each. During the first ramping sequence, atoms in state $m_F=0$ are brought to self-organization, then, the spin state is changed to the spin mixture, and the atoms are brought to self-organization again. The measurement is performed at a detuning $\Delta_{\mathrm{c}}/2\pi = \SI{-15.40(18)}{\mega\hertz}$.

\subsection*{Theory}
\subsubsection*{Time evolution of the density and the spin mode}

The many-body Hamiltonian describing our system is \cite{Landini2018}: 
\bea
\hat{H} &&= \hbar(-\Delta_{\mathrm{c}} - i\kappa)\hat{a}^{\dagger}\hat{a} + \hbar\omega_0 (\hat{J}_{z,+} + \hat{J}_{z,-}) + \frac{\hbar}{\sqrt{N}}\Big[ \lambda_{\mathrm{D}}(\hat{a}^{\dagger} + \hat{a})(\hat{J}_{x,+} + \hat{J}_{x,-}) + i\lambda_{\mathrm{S}} (\hat{a}^{\dagger} - \hat{a})(\hat{J}_{x,+} - \hat{J}_{x,-})\Big] \nonumber \\
&& = \hbar(-\Delta_\mathrm{c} - i\kappa)\hat{a}^{\dagger}\hat{a} + \hbar\omega_0 (\hat{J}_{z,+} + \hat{J}_{z,-}) + \frac{\hbar}{\sqrt{N}}\lambda_1\Big[\hat{J}_{x,+}(\hat{a}^{\dagger}e^{i\phi_1} + \hat{a}e^{-i\phi_1})  + \hat{J}_{x,-}(\hat{a}^{\dagger}e^{-i\phi_1} + \hat{a}e^{i\phi_1})\Big] \quad, 
\label{eq:many body Hamiltonian}
\eea

where $\hat{a}$ ($\hat{a}^{\dagger}$) is the annihilation (creation) operator corresponding to the cavity mode in the frame of the transverse pump frequency. The atomic system is represented by an ensemble of $N$ effective "spins" in each of the two Zeeman states with $\hat{J}_{x,\pm}$, $\hat{J}_{y,\pm}$ and $\hat{J}_{z,\pm}$ being the corresponding angular momentum operators. This pseudo spin is constructed from the macroscopically occupied zero momentum state $| 0 \rangle$ of the BEC and an excited state $| k \rangle$ with a symmetric superposition of four momentum states $|p_x = \pm \hbar k,p_z = \pm \hbar k\rangle$ which represent one recoil momentum $\hbar k$ each in the cavity and the transverse pump direction \cite{Baumann2010}. The wave number of the transverse laser field with wavelength $\lambda$ is given by $k = 2\pi/\lambda$. For zero lattice depth $V_\mathrm{TP}$ of the transverse pump, the energy difference between the states $|0\rangle$ and $|k\rangle$ is $\hbar\omega_0 = \SI{2}{\Er} = \hbar^2 k^2/m$ with $m$ being the mass of a $^{87}$Rb atom and $\hbar$ is the Planck's constant divided by $2\pi$.\\

The occupation of the density and the spin mode results in density and spin modulation in the system which are quantified by the expectation values: $x_{\mathrm{D}} = \frac{1}{N}(\avg{\hat{J}_{x,+}} + \avg{\hat{J}_{x,-}})$ and $x_{\mathrm{S}} = \frac{1}{N}(\avg{\hat{J}_{x,+}} - \avg{\hat{J}_{x,-}})$, respectively. Any finite value of $\avg{\hat{J}_{x,\pm}}$ physically implies a $\lambda$-periodic checkerboard modulation: the density mode consists of spatially identical modulation patterns for both Zeeman states, which are on the other hand relatively shifted by $\lambda/2$ for the spin mode. As the coupling with the density mode is mediated via the real quadrature $(\hat{a}^{\dagger} + \hat{a})$ of the cavity, this quadrature is simultaneously occupied with the density mode. Similarly, the occupation of the spin mode is accompanied with the population of the imaginary quadrature $i(\hat{a}^{\dagger} - \hat{a})$ of the cavity. The damping of the cavity mode is represented by the non-Hermitian term $-i\kappa \hat{a}^{\dagger}\hat{a}$ in Eq. \ref{eq:many body Hamiltonian}. \\

The coupling strengths $\lambda_{\mathrm{D,S}}$ are defined as: $\lambda_\mathrm{D} = M_0E_\mathrm{p}E_0\alpha_\mathrm{s}\sqrt{N}\cos{\varphi}$ and $\lambda_\mathrm{S} = M_0E_\mathrm{p}E_0 (\alpha_\mathrm{v}/2)\sqrt{N}\sin{\varphi}$. Here $E_\mathrm{p} = \sqrt{-4V_{\mathrm{TP}}/\alpha_\mathrm{s}}$ and $E_0$ represent electric field amplitudes of the transverse pump and the empty cavity, respectively. $\alpha_{\mathrm{s,v}}$ represent the scalar and vectorial components of the polarizability tensor with ratio $\alpha_\mathrm{v}/2\alpha_\mathrm{s} = 0.464$ for the operating wavelength $\lambda = \SI{784.7}{\nano\meter}$ of the transverse pump. $M_0 = \langle 0 | \cos{kx}\cos{kz} | k \rangle$ defines an overlap integral where $|0 \rangle $ and $|k \rangle $ are the ground and excited momentum states. The energy difference between the two momentum states and the overlap $M_0$ depends on the lattice depth $V_\mathrm{TP}$ of the standing wave potential created by the transverse pump. We further rewrote the Hamiltonian in Eq. \ref{eq:many body Hamiltonian} by defining $\lambda_1^2 = \lambda_\mathrm{D}^2 + \lambda_\mathrm{S}^2$ and $\tan{\phi_1} = \frac{\lambda_\mathrm{S}}{\lambda_\mathrm{D}}$.\\

By using Heisenberg's equations of motion, we write the time evolution of $\avg{\hat{a}^{\dagger}}$, $\avg{\hat{J}_{x,\pm}}$ and $\avg{\hat{J}_{y,\pm}}$ as: 

\beq
\frac{d}{dt}\avg{ \hat{a}^\dagger} = -(\kappa + i\Delta_\mathrm{c})\avg{\hat{a}^{\dagger}} + i\frac{\lambda_1}{\sqrt{N}}\Big[\avg{\hat{J}_{x,+}} e^{-i\phi_1} + \avg{\hat{J}_{x,-}} e^{i\phi_1}\Big]\quad,
\label{time evolution:cavity-field}
\eeq

\beq
\frac{d}{dt} \avg{\hat{J}_{x,\pm}}  = -\omega_0 \avg{\hat{J}_{y,\pm}} \quad ,
\label{time evolution:Jx}
\eeq

\beq
\frac{d}{dt}\avg{\hat{J}_{y,\pm}}  = \omega_0 \avg{ \hat{J}_{x,\pm}} -\frac{\lambda_1}{\sqrt{N}}\Big[\avg{\hat{a}} e^{\mp i\phi_1} + \avg{\hat{a}^{\dagger}} e^{\pm i\phi_1}\Big] \avg{ \hat{J}_{z,\pm}} \quad.
\label{time evolution:Jy}
\eeq
Here we further employed the factorization \cite{Dimer2007}: $\avg{\hat{a}\hat{J}_{z,\pm}} \rightarrow \avg{\hat{a}}\avg{\hat{J}_{z,\pm}}$ and $\avg{\hat{a}^{\dagger}\hat{J}_{z,\pm}} \rightarrow \avg{\hat{a}^{\dagger}}\avg{\hat{J}_{z,\pm}}$. As the cavity field approaches a steady state at a rate $\kappa/2\pi = \SI{1.25}{\mega\hertz}$ which is much faster than the motional frequency $2\omega_0/2\pi \approx \SI{7.4}{\kilo\hertz}$ of the atoms, we can adiabatically eliminate the cavity field which gives:
\beq
\avg{ \hat{a}^{\dagger}} = -\frac{\lambda_1}{\sqrt{N}}\frac{1}{\sqrt{\Delta_\mathrm{c}^2 + \kappa^2}}e^{-i\phi_{\kappa}}\Big[\langle \hat{J}_{x,+} \rangle e^{-i\phi_1}+ \langle \hat{J}_{x,-} \rangle e^{i\phi_1}\Big]\quad,
\label{cavity-field}
\eeq
where $\phi_{\kappa} = \tan^{-1}\big(-\kappa/\Delta_\mathrm{c}\big)$ is the cavity induced phase shift. By combining Eq. \ref{time evolution:Jx} to \ref{cavity-field}, we obtain the following second order differential equations for $\avg{\hat{J}_{x,\pm}}$:
\beq
\frac{d^2}{dt^2} \avg{\hat{J}_{x,+}} = -\omega_0^2\avg{\hat{J}_{x,+}} -\frac{2\lambda_1^2\omega_0}{\Delta_\mathrm{c}^2 + \kappa^2}\frac{1}{N}\Big[-\Delta_\mathrm{c} \avg{\hat{J}_{x,+}} + (\Delta_\mathrm{c}\sin{\delta\phi} + \kappa \cos{\delta\phi})\avg{\hat{J}_{x,-}} \Big]\avg{\hat{J}_{z,+}}\quad,
\eeq
\beq
\frac{d^2}{dt^2} \avg{\hat{J}_{x,-}} = -\omega_0^2\avg{\hat{J}_{x,-}} -\frac{2\lambda_1^2\omega_0}{\Delta_\mathrm{c}^2 + \kappa^2}\frac{1}{N}\Big[-\Delta_\mathrm{c} \avg{\hat{J}_{x,-}} + (\Delta_\mathrm{c}\sin{\delta\phi} - \kappa \cos{\delta\phi})\avg{\hat{J}_{x,+}} \Big]\avg{\hat{J}_{z,-}}\quad,
\eeq
where $\delta \phi = 2\phi_1 - \pi/2$. By assuming only small deviations from the initial ground state, we can approximate $\avg{\hat{J}_{z,\pm}} \approx -\frac{N}{2}$. With the definitions $x_\mathrm{D} = \frac{1}{N}(\avg{\hat{J}_{x,+}} + \avg{\hat{J}_{x,-}})$ and $x_\mathrm{S} = \frac{1}{N}(\avg{\hat{J}_{x,+}} - \avg{\hat{J}_{x,-}})$, we obtain equation (2) of the main text which is reiterated below: 
\beq
\frac{d^2}{dt^2}
\left(
\begin{array}{c}
	x_\mathrm{D}\\
	x_\mathrm{S}\\
\end{array}
\right)
= \left(
\begin{array}{cc}
	-\omega_\mathrm{D}^2 & -K^2 \\
	K^2 & -\omega_\mathrm{S}^2 \\
\end{array}
\right)
\left(
\begin{array}{c}
	x_\mathrm{D}\\
	x_\mathrm{S}\\
\end{array}
\right)\quad,
\label{eq:dissipatively coupled modes SI}
\eeq
where the soft mode frequencies $\omega_\mathrm{D,S}$ and the strength $K^2$ of the dissipative coupling are defined via the following equations:
\[
\omega_\mathrm{D}^2 = \omega_0^2 + \frac{\lambda_1^2 \omega_0\Delta_\mathrm{c}}{\Delta_\mathrm{c}^2 + \kappa^2}\Big[ 1 - \sin{\delta\phi}\Big]\quad,
\]
\[
\omega_\mathrm{S}^2 = \omega_0^2 + \frac{\lambda_1^2 \omega_0\Delta_\mathrm{c}}{\Delta_\mathrm{c}^2 + \kappa^2}\Big[ 1 + \sin{\delta\phi}\Big]\quad,
\]
\[
K^2 = \frac{\lambda_1^2\omega_0\kappa}{\Delta_\mathrm{c}^2 + \kappa^2}\cos{\delta\phi} = \frac{1}{\kappa}{\lambda_1^2\omega_0\sin^2{\phi_{\kappa}}}\cos{\delta\phi}\quad.
\]

As seen from above relations, the cavity decay rate leads to a small shift in the soft mode frequencies which slightly modifies the critical transition point for the occupation of the density and the spin mode (in the absence of the off-diagonal dissipative coupling). This is the reason why we consider the coupling between the spinor BEC and the DM or the SM mode to be coherent. As shown for the case of single spin BEC \cite{Brennecke2013}, dissipation might also alter the critical exponent for the self-organization phase transition corresponding to the occupation of the density or the spin mode.

\subsubsection*{Description of the dissipation induced instability}
To describe the response of the system to the dissipative force, we need to solve  Eq. \ref{eq:dissipatively coupled modes SI}. Assuming solutions of the form $[x_\mathrm{D},x_\mathrm{S}] = [c_\mathrm{D},c_\mathrm{S}] e^{\eta t}$, we obtain $\eta$ to be:
\beq
\eta^2 = -\frac{\omega_\mathrm{D}^2 + \omega_\mathrm{S}^2}{2} \pm \frac{1}{2} \sqrt{(\omega_\mathrm{D}^2 - \omega_\mathrm{S}^2)^2 - 4K^4}\quad.
\label{eq:eta}
\eeq
A dynamically unstable chiral solution requires Re\{$\eta$\} to be positive which corresponds to amplification and Im\{$\eta$\} to be non-zero. These conditions are satisfied when $(\omega_\mathrm{D}^2 - \omega_\mathrm{S}^2)^2 - 4K^4 < 0$ which results in $\tan{|\delta \phi|} < -\frac{\kappa}{\Delta_\mathrm{c}}$ for the dynamically unstable regime. Note that outside the unstable region, $\eta$ is either real or imaginary corresponding to normal state or static self-organized states, respectively.\\

The exact solutions of $\eta$ as obtained from Eq. \ref{eq:eta} and including the effects of cavity birefringence and dispersive shift (see next subsection) are plotted in Fig. \ref{fig:Fig3}B of the main text. In the limit $|\omega_\mathrm{D}^2 - \omega_\mathrm{S}^2| << K^2 << |\omega_\mathrm{D}^2 + \omega_\mathrm{S}^2|$, we obtain the following four solutions for $\eta$:
\[
\eta \approx \pm i\sqrt{\frac{\omega_\mathrm{D}^2 + \omega_\mathrm{S}^2}{2}}\Big[1 \mp i\frac{K^2}{\omega_\mathrm{D}^2 + \omega_\mathrm{S}^2} \Big] \quad,
\]
whose imaginary and real parts are the approximate expressions of the rotation frequency $\omega$ and the amplification rate $g$ as given in the main text. We also observe that when $\omega_\mathrm{D,S}^2 < 0$ (which would correspond to the self-organization phase transition in the absence of dissipation), the real and imaginary parts of $\eta$ are interchanged. Note that there are two solutions of $\eta$ which are amplified and two which are damped. Using the two amplified solutions of $\eta$, we construct the time varying solutions $x^\mathrm{c}_\mathrm{D,S}$ of the occupation of the density and spin mode:
\[
\big[x_\mathrm{D}^\mathrm{c},x_\mathrm{S}^\mathrm{c}\big] = A\big[ \cos{\omega t},\sin{(\omega t + \frac{\Delta_\mathrm{c}}{\kappa}\delta \phi)} \big] e^{gt}\quad,
\]
where $A$ is the amplitude of the initial fluctuation leading to the instability. We have, for simplicity, skipped an additional phase offset in the arguments of $\sin$ and $\cos$ functions above which describes the density and the spin admixture of the initial fluctuation. The axis of the elliptical solution can be more easily interpreted by a $45^{\circ}$ rotation of the $x_\mathrm{D}-x_\mathrm{S}$ basis and defining $x_{\pm} = (x_\mathrm{D} \pm x_\mathrm{S})/2$. Physically, $x_{\pm} = \avg{\hat{J}_{x,\pm}}/N$ corresponds to the normalized strength of the checkerboard modulation for each of the two spin states. These solutions are given by:
\[
\big[x_+^\mathrm{c},x_-^\mathrm{c}\big] = A\big[ \cos{(\phi_2/2)}\cos{(\omega t + \phi_2/2)},\sin{(\phi_2/2)}\sin{(\omega t  + \phi_2/2)} \big] e^{gt}\quad,
\]
where $\phi_2 = -\frac{\pi}{2} + \frac{\Delta_\mathrm{c}}{\kappa}\delta \phi$. This shows that the elliptical solutions are oriented in the $x_{\pm}$ directions as plotted in Fig. \ref{fig:Fig2}. Finally, we derive an expression for the cavity-output field. Following Eq. \ref{cavity-field}, we get
\[
\avg{\hat{a}^{\dagger}} \propto x^\mathrm{c}_{+} e^{-i\phi_1}+ x^\mathrm{c}_{-} e^{i\phi_1} \propto (x_\mathrm{D}^\mathrm{c} - ix_\mathrm{S}^\mathrm{c}) -\frac{\delta \phi}{2} (x_\mathrm{D}^\mathrm{c} + i x_\mathrm{S}^\mathrm{c}) = \Big[A_+{e^{i\omega t}} + A_-{e^{-i\omega t}} \Big]e^{gt}\quad,
\]
where the last proportionality also assumes small deviations from the critical polarization angle and one can show that $\frac{A_+}{A_-} \approx -\frac{\delta \phi}{2}(1 + i\frac{\Delta_\mathrm{c}}{\kappa})$. 

\subsubsection*{Adding the effects of cavity birefringence and dispersive shift}
In the experiment, the cavity has two birefringent $\mathrm{TEM}_{00}$ modes whose polarization axes are tilted by $\alpha = 22^{\circ}$ with respect to the y and z axes. The spinor BEC is coupled to the density and the spin mode via both these cavity modes. The detuning $\Delta_{\mathrm{c}}$ mentioned in the main text is associated to the resonance of the mainly y-polarized mode whose frequency is $\delta_{\mathrm{bir}}/2\pi = \SI{2.2}{\mega\hertz}$ larger than the other cavity mode. The effect of coupling to these two cavity modes can be taken into account in the definition of $\omega_\mathrm{D,S}$ and $K^2$ as follows:
\[
\frac{\Delta_{\mathrm{c}}}{\Delta_{\mathrm{c}}^2 + \kappa^2} \rightarrow \frac{\Delta_{\mathrm{c}}}{\Delta_{\mathrm{c}}^2 + \kappa^2}\cos^2{\alpha} + \frac{\Delta_{\mathrm{c}} + \delta_{\mathrm{bir}}}{(\Delta_{\mathrm{c}} + \delta_{\mathrm{bir}})^2 + \kappa^2}\sin^2{\alpha} 
\]
\[
\frac{\kappa}{\Delta_{\mathrm{c}}^2 + \kappa^2} \rightarrow \frac{\kappa}{\Delta_{\mathrm{c}}^2 + \kappa^2}\cos^2{\alpha} + \frac{\kappa}{(\Delta_{\mathrm{c}} + \delta_{\mathrm{bir}})^2 + \kappa^2}\sin^2{\alpha}
\quad .
\]

We further add the effect of the atomic dispersive shift on the cavity resonance and hence detuning $\Delta_{\mathrm{c}}$ in Fig. \ref{fig:Fig2} and \ref{fig:Fig3} of the main text and Fig. \ref{fig:FigSI2} and \ref{fig:Fig4}. The dispersively shifted detuning $\tilde{\Delta}_{\mathrm{c}}$ is defined as: $\tilde{\Delta}_{\mathrm{c}} = \Delta_{\mathrm{c}} - U_0(N_{+1} + N_{-1})\avg{\cos^2 kx}$ where $U_0/2\pi = \SI{-56.3}{\hertz}$ is the maximum dispersive shift per atom and $N_{\pm 1}$ is the atom number in each of the two Zeeman states. $\avg{\cos^2 kx}$ is the overlap of the cavity mode with the atomic wavefunction and is equal to 1/2 for the non-modulated BEC. The light field scattered into the cavity changes this overlap. We neglect this dynamical shift while plotting $\tilde{\Delta}_{\mathrm{c}}$ as the maximum number of photons in the cavity is $\approx 100$. This leads to an intra-cavity lattice depth of $\approx \SI{1.5}{\Er}$ which results in only a small dynamical shift that is irrelevant for this study. 

\subsubsection*{Experimental justification of the model}
An important approximation employed in our derivation of the equations of motion is: $\avg{\hat{J}_{z,\pm}} \approx -\frac{N}{2}$. This can be verified experimentally as we can use the obtained photon number and Eq. \ref{cavity-field} to extract $\avg{\hat{J}_{x,\pm}}$. For the instability data shown in Fig. \ref{fig:Fig2}, this gives an estimated change in the value of $\avg{\hat{J}_{z,\pm}}$ of at most 23\% for a binsize of \SI{5}{\milli\s}. Also, note that the reason why the change in $\avg{\hat{J}_{z,\pm}}$ should be small is not built-in to our simple non-interacting model and we attribute it to the s-wave collisional interactions, see below. We also converted the number of photons in steady state self-organization to $\avg{\hat{J}_{x,\pm}}$ (data for Fig. \ref{fig:Fig2}A). From the non-interacting theory, $|\avg{\hat{J}_{x,\pm}}| < \frac{N}{2}$, but we found the corresponding experimental values to be as large as $N$ which cannot be fully captured within our calibration errors.

\subsection*{Data Processing}
\subsubsection*{Spectrum fitting}
\begin{figure}
	\centering
	\includegraphics[keepaspectratio=true,]{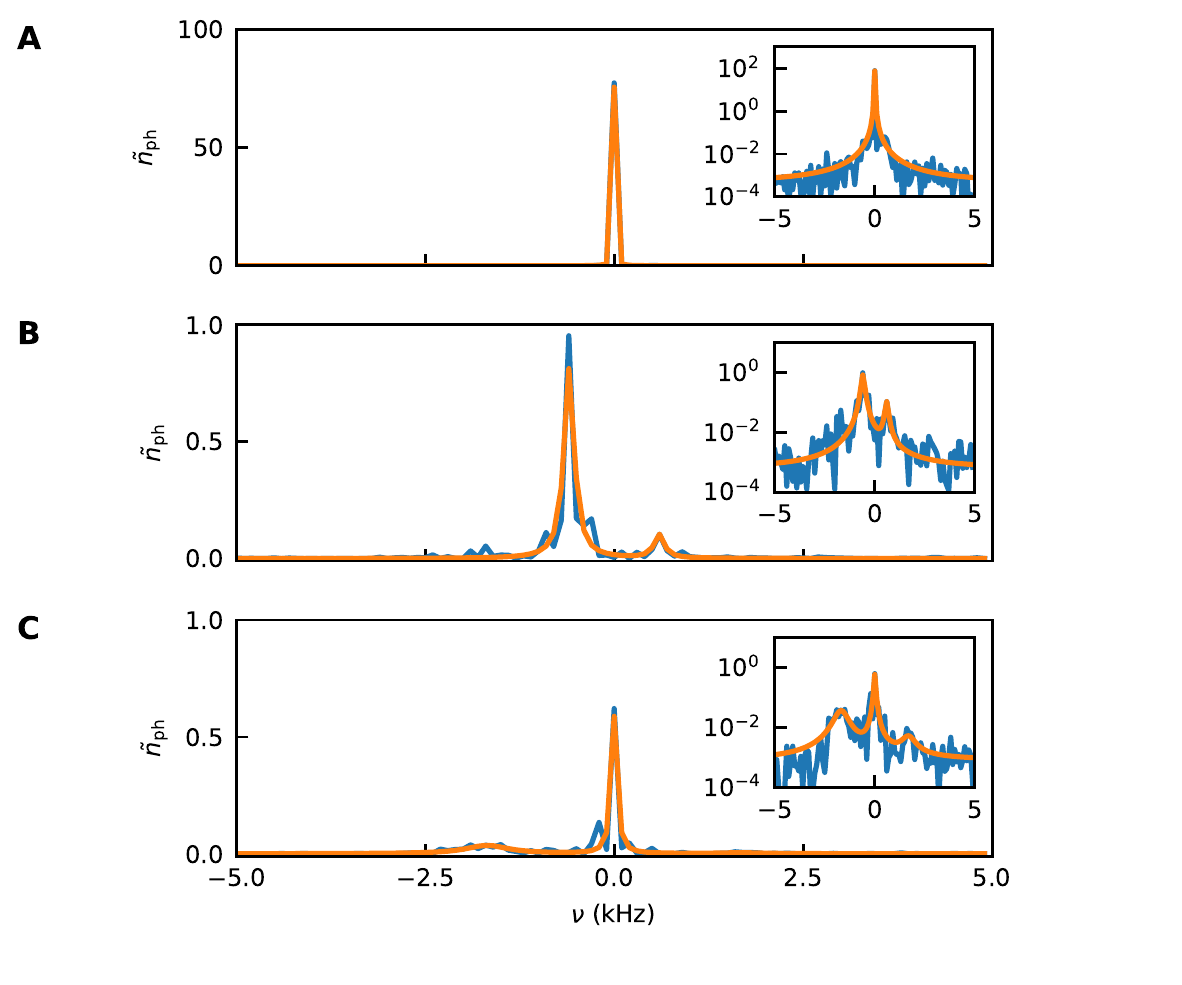}
	\caption{Exemplary fitted $\tilde{n}_\mathrm{ph}$ for a detuning $\Delta_\mathrm{c}/2\pi$ of -9.2(5) MHz and different transverse pump powers $V_\mathrm{TP}$. (\textbf{A}) $V_\mathrm{TP}=25.0(6) E_\mathrm{R}$, dominant peak at zero frequency, corresponding to a self-organized state. (\textbf{B}) $V_\mathrm{TP}=21.3(6) E_\mathrm{R}$, PSD in the unstable region, characterized by the two sidebands. (\textbf{C}) $V_\mathrm{TP}=11.4(6) E_\mathrm{R}$, intermediate situation with all three components being present. We assign these instances to self-organization, due to the larger weight on the central peak. The data was recorded with an acquisition time $T$ of 10 ms. Insets show the same data on logarithmic scale.}
	\label{fig:FigSI1}
\end{figure}

We record the time trace of the cavity emission by means of a heterodyne detector \cite{Baumann2011}. The light field leaking from the cavity is mixed with a local oscillator, which is frequency shifted by 60 MHz with respect to the transverse pump frequency, on a balanced heterodyne detector. The signal is down-converted with an RF mixer to 47kHz and recorded with an analog-to-digital converter with a time resolution of \SI{2}{\micro\second}. Heterodyning allows us to access both amplitude and phase information of the cavity light field. The two light quadratures $X(t),Y(t)$, after calibration, are combined to give the cavity field $\alpha^*(t)=X+ i Y$. After performing digitally a rotation at \SI{47}{\kilo\hertz}, we perform fast Fourier transform (FFT) of the signal in a given time window $T$ to access the spectrum $\tilde{\alpha}^*(\nu)=1/\sqrt{N}\sum_i \alpha^*(t_i)e^{-i 2\pi\nu t_i}$, where $t_i$ is the time step and N is the total number of time steps contained in the time window. From this we obtain the power spectral density as: $PSD(\nu)=\tilde{\alpha}^*(\nu) \tilde{\alpha}(\nu)=\tilde{n}_\mathrm{ph}(\nu)T$.
We fit the power spectral density of the cavity emission with an empirical model consisting of three Lorentzian peaks. 
\begin{equation}
\tilde{n}_\mathrm{ph}(\nu)=\frac{a}{1+(\nu-\nu_0)^2/\Gamma^2}+\frac{b}{1+(\nu+\nu_0)^2/\Gamma^2}+\frac{c}{1+\nu^2/\Gamma_0^2}\quad ,
\end{equation}
where $a,b,c,\nu_0,\Gamma,\Gamma_0$ are fitting parameters (see Fig. \ref{fig:FigSI1}).
The amplitude of the zero frequency component $c$ is connected to the value of the order parameter $\langle \hat{J}_x \rangle$ in steady state self-organization. The other two amplitudes $a,b$ are signaling the onset of the positional instability, with a mean frequency $\nu_0$. The value of $\Gamma_0$ is typically at or below the Fourier limit, while $\Gamma$ can take values ranging from the Fourier limit to $\Gamma\simeq\nu_0$, depending on the experimental parameters. Typically, for regions of parameters close to the onset of the instability, we observe low values of $\Gamma$, while the values increase deeper into the unstable region. Whenever the integrated signal is below the noise level, we consider the system to be in the normal phase. Otherwise, if the integrated signal from the central peak is above the integrated signal of the side peaks, we consider the system to be self-organized. Vice-versa we consider the system to be dominated by the instability. This logic excludes the possibility for the two orders to coexist. Even though this might be the case in some regions of parameters, the understanding of the system in such conditions goes beyond the scope of this publication. Typical values for the photon number $n_\mathrm{ph}=\sum_{\nu} \tilde{n}_\mathrm{ph}$ in the self-organized phase are 10-100 photons, while in the unstable region the signal is about 1-10 photons. The background noise level is around $10^{-2}$ photons.

\subsubsection*{Extracting the amplitude and phase of the cavity light field}
To obtain the number of photons $n_{\mathrm{ph}}$ and phase $\phi$ as shown in Fig. \ref{fig:Fig2}A-B, we digitally rotate the heterodyne signal at 47 kHz to obtain the light field at the frequency of the transverse pump. We further remove high frequency noise from the signal by performing a moving average with a 100 $\mu$s window.

\subsubsection*{Construction of phase-space trajectories}
For the sub-panels displaying the $x_{\mathrm{D}} –x_{\mathrm{S}}$-plane of Fig. \ref{fig:Fig2} of the main text, we recorded the time traces $(X(t),Y(t))$ over 20 runs of the experiment under the same experimental conditions. In order to meaningfully average the data we have to take particular consideration of the phase of the oscillation. Due to technical imperfections, the phase reference of the heterodyne detection is uncorrelated between different experimental runs. This effectively rotates the trajectories relative to each other by a random amount in every experimental repetition. To correct for this effect, we use two different procedures, depending on the system being either in the self-organized region or in the instability region. For data in the self-organized region, we just evaluate the phase $\phi_0$ of $\tilde{\alpha}^*(\nu=0)$, and phase shift the spectrum by $-\phi_0$. For data in the instability dominated region, we evaluate the phase of $\tilde{\alpha}^*(\nu)$ at the frequency corresponding to the maximum of $\tilde{n}_\mathrm{ph}$ on the negative side of the frequency spectrum ($\phi_-$) and at the corresponding frequency on the positive side ($\phi_+$). We therefore phase shift $\tilde{\alpha}^*$ by $-(\phi_++\phi_-)/2$ and invert the FFT to generate the corrected quadratures $X'(t),Y'(t)$. Looking for the contribution at the frequency of the maximum, we have
\begin{eqnarray}
X'(t)&=&(|\tilde{\alpha}|_++|\tilde{\alpha}|_-)\cos(2\pi\nu t+\delta),\\
Y'(t)&=&(|\tilde{\alpha}|_+-|\tilde{\alpha}|_-)\sin(2\pi\nu t+\delta),
\end{eqnarray}
where $|\tilde{\alpha}|_{+,-}$ are the amplitudes at the two frequencies and $\delta=(\phi_+-\phi_-)/2$. With this procedure, we get an ellipse oriented along the x,y-axes. We can therefore meaningfully average the $(X'(t),Y'(t))$ traces. This procedure renders impossible a determination of the effective orientation of the ellipses which would provide information on the eigenmodes of the system. Such information could be accessed by generating a phase reference in each shot as done in \cite{Baumann2011}. The dataset displays a significant modulation of the oscillation amplitude over time as visible on the non-averaged trace in Fig. \ref{fig:Fig3}B of the main text. The elliptical orbit is correspondingly showing a modulation of the mean radius, if considering data from the full time trace. Therefore, the data segment plotted in the insets corresponds to 500 $\mu$s of evolution around the first recorded peak. The data is furthermore smoothened with a moving average of 40 $\mu$s to reduce high frequency noise.

\subsubsection*{Determining the boundary of the instability dominated region}
\begin{figure}
	\centering
	\includegraphics[keepaspectratio=true,]{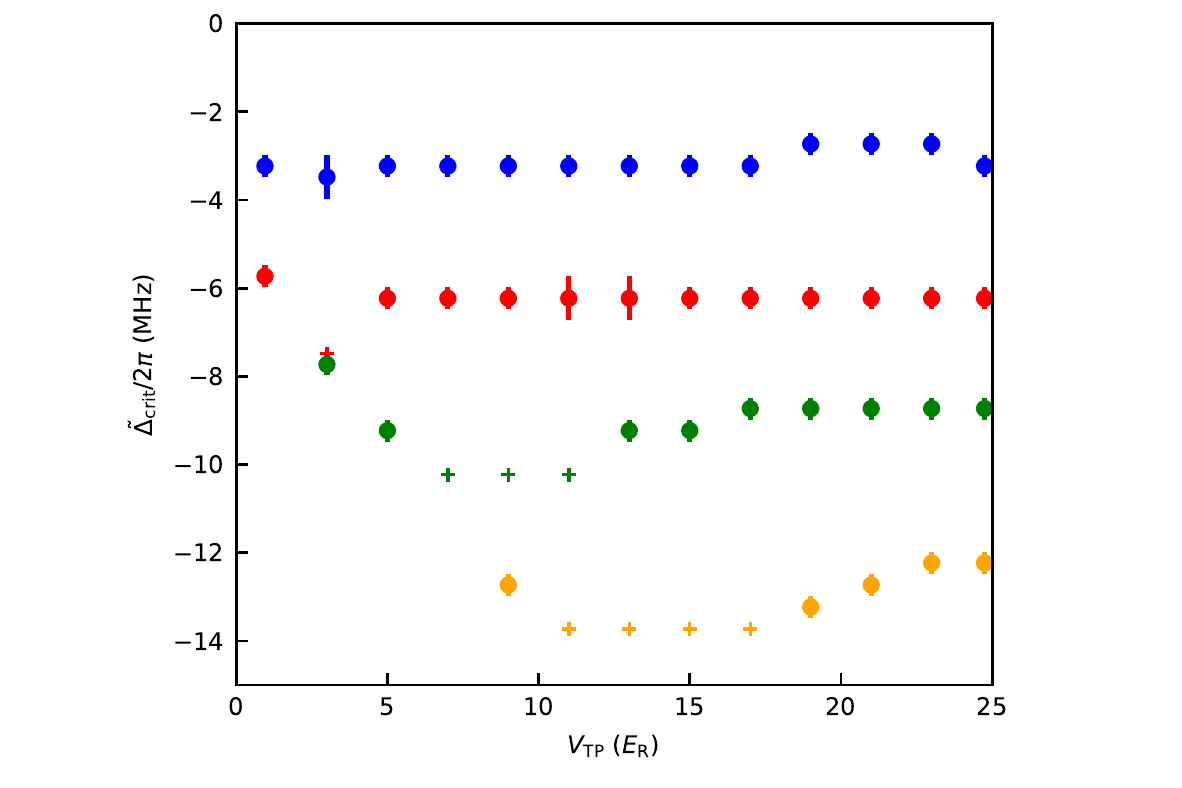}
	\caption{Exemplary determination of the dispersively shifted critical detuning $\tilde{\Delta}_\mathrm{crit}$ defining the boundary between self-organization and instability dominated regime as a function of $V_\mathrm{TP}$ for different polarization angles $\varphi=3.4(1.0)^\circ$ (blue), $\varphi=54.4(1.0)^\circ$ (red), $\varphi=61.4(1.0)^\circ$ (green), $\varphi=62.9(1.0)^\circ$ (orange). A weak power dependence can be noticed. The data points in Fig. 3 (main text) are measured at $V_\mathrm{TP}=21 E_\mathrm{R}$. The experimentally determined cavity detunings are binned with a homogeneous bin size of 0.5 MHz. The error bar is half of the bin size, unless the data set has insufficient statistical relevance around the transition point. In such a case the error bar is the full bin size. Plus signs represent instances in which a precise determination of the boundary was not possible due to the finite detuning range employed in the measurement. The position of such symbols gives an upper bound for the critical detuning.}
	\label{fig:FigSI2}
\end{figure}

In Fig. \ref{fig:Fig3} of the main text we plot the region of the dissipation induced instability as a function of detuning and polarization angle for a lattice depth of \SI{21}{\Er}. The transverse pump power could in principle contribute to the position of the boundary as well. In order to obtain the data reported, we measured the spectrum for fixed polarization angle and detuning as a function of time while linearly ramping up the transverse pump power to the maximum value of 25 $E_\mathrm{R}$ in 50 ms. The time trace was segmented in time intervals of 5 ms with a 50\% overlap between neighboring intervals. For each interval we recorded $\tilde{n}_\mathrm{ph}$ and fitted it with our empirical model. We observed that the boundary $\tilde{\Delta}_{\mathrm{crit}}$ between the instability dominated region and steady state self-organization depends very weakly on pump power with significant deviations only close to the boundary with the normal phase, see Fig. \ref{fig:FigSI2}. This is in line with the non-interacting theory which predicts no dependence of $\tilde{\Delta}_{\mathrm{crit}}$ on the lattice depth.

\subsubsection*{Constructing the phase diagram}
\begin{figure}
	\centering
	\includegraphics[keepaspectratio=true,]{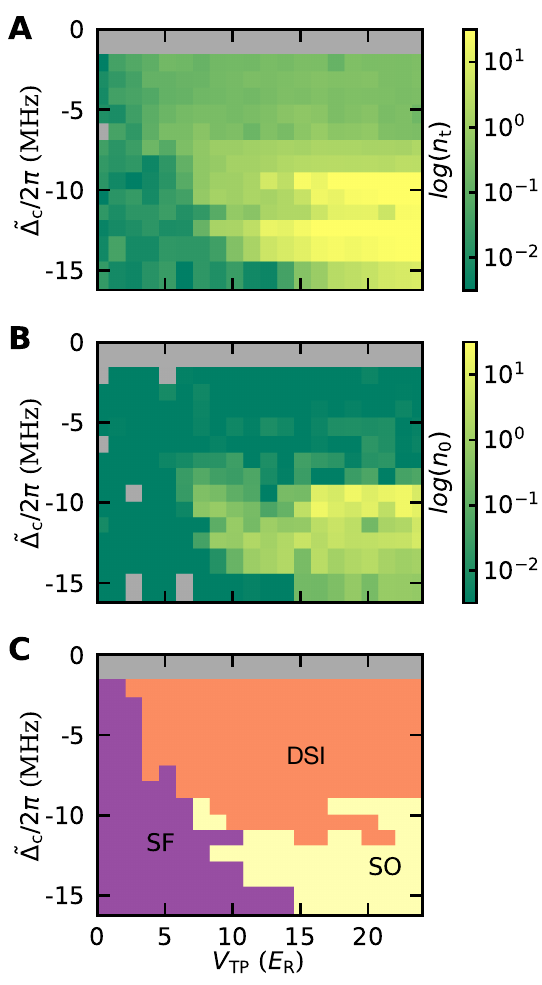}%
	\caption{Constructing a phase diagram: (\textbf{A}) Total number of mean intra-cavity photons $n_{\mathrm{t}}$ and (\textbf{B}) number of mean intra-cavity photons $n_{\mathrm{0}}$ at the frequency of the transverse pump as a function of dispersively shifted detuning $\tilde{\Delta}_{\mathrm{c}}$ and lattice depth $V_{\mathrm{TP}}$ of the transverse pump. The photon numbers are obtained from the spectrograms using a $\SI{10}{\milli\s}$ window while holding all external parameters constant. (\textbf{C}) Using $n_{\mathrm{t}}$ and $n_{\mathrm{0}}$, we construct a phase diagram depicting two time independent states: superfluid state SF (purple) with negligible mean intra-cavity photon number, and self-organized state SO (yellow) where the dominant contribution of the cavity field is at the frequency of the transverse pump. The dissipation induced structural instability DSI where the state of the system is varying in time and the dominant contribution to the cavity field are in the sidebands is depicted in orange. The measurement is performed at a polarization angle $\varphi = 63.3(1.0)^{\circ}$ and every data point is obtained by averaging over 5 experimental repetitions.}
	\label{fig:Fig4}
\end{figure}

In Fig. \ref{fig:Fig4}, we construct the phase diagram of the system as a function of lattice depth and detuning at $\varphi= 63.3(1.0)^\circ$. The transverse pump power is ramped up via the S-shaped ramp in \SI{20}{\milli\second} to a lattice depth ranging between 0 to \SI{20}{\Er}, held for \SI{50}{\milli\second}, and ramped down via the S-shaped ramp in \SI{20}{\milli\second}. The power spectral density and hence $\tilde{n}_\mathrm{ph}$ is extracted from the first \SI{10}{\milli\second} after the end of the upwards ramp. This spectrum is fitted with our empirical model to extract two order parameters: the total mean number of photons $n_{\mathrm{t}}$ and the mean number of photons $n_0$ at the frequency of the transverse pump, Fig. \ref{fig:Fig4}A-B. Based on these order parameters, we build a phase diagram exhibiting three regimes as shown in Fig. \ref{fig:Fig4}C. The superfluid state SF is characterized by negligible intra-cavity light field. In contrast, the self-organized state SO \cite{Baumann2010} which corresponds to the occupation of the DM is identified by cavity light field dominantly at the transverse pump frequency, $n_{\mathrm{t}} \approx n_0 \neq 0$. Finally, the dissipation induced structural instability leads to a finite occupation of the red and blue sidebands while there is negligible power at the frequency of the transverse pump, $n_{\mathrm{t}} \neq 0 \approx n_0$.

\subsubsection*{Adiabaticity and heating}
The lifetime of the system in the unstable region was observed to be around 5-10 ms, rendering impractical very long ramp or observation time scales. The choice of time scales for the power ramps represents a compromise between adiabaticity and heating. 

\subsubsection*{Critical polarization angle determination}
We determined the value of the critical polarization angle with two different methods. The first one relies on fitting the data from Fig. \ref{fig:Fig3}A to determine the position of the minimum. This method gives the value $63.7(1.0)^\circ$, where the error bar is dominated by the repeatability of our polarization setting $\simeq1^\circ$. For the second method we looked at the phase of the light during self-organization for large detuning ($\Delta_{\mathrm{c}}/2\pi = \SI{-15.04(18)}{\mega\hertz}$) from cavity resonance using the same experimental procedure as in \cite{Landini2018}. We observe a clear phase jump of $\pi/2$. Fitting the position of the phase jump with a sigmoid function yields the value $62.2(1.0)^\circ$, where again the uncertainty is dominated by the repeatability of the polarization setting. In this measurement, we also observe that for angles around the critical angle, the observed value of the phase goes smoothly from zero to $\pi/2$ in a window of $\approx 3^{\circ}$. The expected value for the critical angle for non-interacting atoms is $65.1^\circ$, deviations from this value are nevertheless expected due to atom-atom interactions disfavoring density modulations over spin-texture, leading to a shift towards lower angles. The second method relies on data taken at higher transverse pump powers (mean densities) with respect to the first method due to the larger cavity detuning. For the theoretical plots presented in the main text, we assumed the value $65.1^\circ$ obtained from the theoretical values of scalar and vectorial polarizabilities at the transverse pump wavelength of $\SI{784.7}{\nano\meter}$ \cite{Landini2018}.

\subsubsection*{Role of collisional interactions}
We observed a much lower mean intra-cavity photon number in the DSI (Fig. \ref{fig:Fig2}D-E) than during self-organization (Fig. \ref{fig:Fig2}C). Such a behavior can be qualitatively explained by the presence of s-wave collisional interactions between the atoms. As the atomic field evolves from one checkerboard pattern to another, these interactions can scatter atoms out of the two considered momentum states. Such processes can limit the amplitude growth in the dissipation induced instability.

From Fig. \ref{fig:Fig4}C, we observe that there is a finite $V_{\mathrm{TP}}$ threshold for the onset of the dissipation induced instability which increases with decreasing detuning. We attribute this finite threshold to a balance between the amplification rate $g$ and a damping rate $\gamma$ due to the collisional interactions \cite{Brennecke2013}. As $g$ decreases with the detuning becoming more negative because of the reduced cavity phase shift, the threshold correspondingly increases. High thresholds for more negative detuning also increase the effect of atomic interactions as the atoms get more and more squeezed in the pancakes formed by the transverse pump lattice. This effect can be responsible for a finite critical detuning for the onset of the dissipation induced instability at the critical angle as opposed to the predictions of the non-interacting theory, cf. Fig. \ref{fig:Fig3}A. However, we also cannot rule out the role of polarization impurity in making this threshold finite.

%

\end{document}